\documentclass[a4paper]{JHEP3}
\usepackage{amsmath,amssymb}
\usepackage{epsfig,cite}

\newcommand{\be}{\begin{equation}}
\newcommand{\ee}{\end{equation}}
\newcommand{\bea}{\begin{eqnarray}}
\newcommand{\eea}{\end{eqnarray}}
\newcommand{\ba}{\begin{array}}
\newcommand{\ea}{\end{array}}

\newcommand{\mrm}[1]{\mathrm{#1}}
\def\fm{\,{\rm fm}}

\def\rmO{\textrm{O}}
\newcommand{\nf}{N_{\rm f}}
\newcommand{\cA}{c_{\rm A}}
\newcommand{\ca}{c_{\rm A}}
\newcommand{\csw}{c_{\rm sw}}
\newcommand{\za}{Z_{\rm A}}
\newcommand{\eq}[1]{eq.~(\ref{#1})}
\newcommand{\fig}[1]{Fig.~\ref{#1}}
\newcommand{\tab}[1]{Table~\ref{#1}}

\newcommand{\psibar}{\bar\psi}
\newcommand{\dirac}[1]{\gamma_#1}
\newcommand{\fa}{f_{\rm A}}
\newcommand{\fp}{f_{\rm P}}
\newcommand{\op}{\mathcal O}
\newcommand{\vecy}{{\bf y}}
\newcommand{\vecx}{{\bf x}}

\newcommand{\zetabar}{\overline{\zeta}}
\newcommand{\rme}{{\rm e}}
\newcommand{\half}{{\textstyle \frac12}}

\title{Non--perturbative improvement of the axial current
for dynamical Wilson fermions}

\author{Michele Della Morte and Roland Hoffmann\\
Institut f\"ur Physik, Humboldt Universit\"at,
Newtonstr. 15, 12489 Berlin, Germany\\
E-Mail: \email{dellamor@physik.hu-berlin.de}, \email{roland@physik.hu-berlin.de}}

\author{Rainer Sommer\\
DESY Zeuthen,
Platanenallee 6, 15738 Zeuthen, Germany\\
E-Mail: \email{rainer.sommer@desy.de}}

\preprint{HU-EP-05/08\\ SFB/CPP-05-07\\ DESY 05-026}

\abstract{
  A non--perturbative determination of the axial current
  improvement coefficient $\cA$ is performed with two
  flavors of dynamical improved Wilson fermions and plaquette
  gauge action. 
  The improvement condition is formulated with
  Schr\"odinger functional boundary conditions and 
  enforced at constant physical volume. Large
  sensitivity is obtained by using two different
  pseudo--scalar states in the PCAC relation.
  We estimate the resulting correction to $F_{\rm PS}$ at
  $\beta=5.2$ to be around $10\%$.
}


\begin{document}

\section{Introduction \label{intro}}

The approach of lattice observables to their
continuum limit can be understood in terms of Symanzik's effective
low energy theory \cite{impr:Sym1,impr:Sym2}. When
applied to QCD with Wilson fermions~\cite{Wilson:1974sk}, 
close to the continuum limit, it predicts that 
scaling violations are dominated by terms 
linear in the lattice spacing $a$. These effects can
be removed by adding a single term to the action,
with properly tuned coefficient $\csw$ \cite{Sheikholeslami:1985ij}.
Linear effects in the lattice spacing can also be systematically
removed in (on-shell) matrix elements of composite operators
\cite{Luscher:1996sc,Luscher:1996vw}. In particular, $\rmO(a)$-improvement
of the axial current requires to add one dimension four
operator, with coefficient $\ca$ (see \eq{impr}, below).
In the quenched approximation, the coefficients $\csw$ and $\ca$ have
been determined non-perturbatively in the relevant range of bare 
coupling (or lattice spacings) in \cite{Luscher:1996ug}. The improvement
conditions, which determine the improvement coefficients, were
derived from the chiral symmetry of the continuum limit.
More precisely, the PCAC relation was required to hold at
finite lattice spacing~\cite{Luscher:1996sc,Luscher:1996vw,Luscher:1996ug}.
As will be detailed in sect.~2.2, the PCAC-relation can be 
considered with different external states.
In \cite{Luscher:1996sc,Luscher:1996vw,Luscher:1996ug}, finite volume states
were chosen, formulated in the framework of
the Schr\"odinger functional. Later, $\ca$ was 
also estimated from the PCAC relation in large volume
\cite{Bhattacharya:2000pn,impr:ca_ukqcd} at a couple of values
of the lattice spacing. Around $a\approx0.1$~fm, the results for $\ca$ obtained
from the finite
volume definition \cite{Luscher:1996ug} differ quite significantly
from those obtained in large volume in \cite{Bhattacharya:2000pn,impr:ca_ukqcd}.
At smaller lattice spacing, the difference decreases.

For the interpretation of this difference,
one should keep in mind that beyond perturbation theory
the improvement coefficients themselves are affected by $\rmO(a)$
ambiguities. In some detail this has been discussed and demonstrated
numerically in \cite{Guagnelli:2000jw}. The  $\rmO(a)$
ambiguity simply corresponds
to the fact that the improved theory is treated up to 
$\rmO(a^2)$ effects.
While this forbids a unique definition of the improved theory,
the $\rmO(a)$ ambiguities can be made to disappear smoothly
if the improvement condition is evaluated with all physical scales kept
fixed, e.g. in units of $r_0\approx0.5\fm$ \cite{Sommer:1993ce}, 
while only the lattice spacing is varied \cite{Guagnelli:2000jw}. We call this the constant 
physics condition. 
At the same time one has to take care that the improvement conditions are 
imposed using low energy states with $E \ll a^{-1}$ since $\rmO(a)$-improvement 
is valid for those only. So far, the methods of
 \cite{Bhattacharya:2000pn,impr:ca_ukqcd}
have not yet been implemented such as to satisfy these conditions. 

In \cite{Durr:2003nc}, two improvement conditions for $\ca$ were studied,
which are easily generalized to respect the above criteria.
They are formulated in finite volume
with Schr\"odinger functional boundary conditions, which furthermore 
helps to render  the numerical evaluation feasible
in full QCD. We will discuss the  improvement conditions 
briefly in sect.~2.2 and choose one of them to compute $\ca$ in the 
$\nf\!=\!2$ theory, where $\csw$ is known from \cite{Jansen:1998mx,Yamada:2004ja}.
The knowledge of $\ca$ is crucial in order to be able to determine the 
pseudoscalar decay constants, but also in order to compute renormalized
quark masses starting from the PCAC masses, as has been done e.g. 
in \cite{Capitani:1998mq,Garden:1999fg}.

\section{Strategy and techniques}
\label{sect_two}

Before going into the details of our strategy and
techniques let us comment again on the constant 
physics condition. In the Schr\"odinger functional 
and neglecting the choice of the quark mass for a moment, the relevant point
is the following.
We need to know how the lattice spacing depends on $\beta$
in order to determine the latter such that a certain $L/a$ corresponds to a 
prescribed value of $L/r_0$. 
This has to be enforced only with
a rather moderate precision, since (sticking with $r_0$ as the reference scale) a
relative error $\Delta$ in the estimate of $r_0/a$ translates
into an error of the improvement constant
which is proportional to $a/L\! \times\! \Delta$. Thus even
if $\Delta$ varies a bit in the considered range of 
lattice spacings, this is quite irrelevant, in particular if 
$\Delta$ is  a smooth function of the lattice spacing.

In the remainder of this section we will state in more detail
how the constant physics condition  is implemented. 
We will also discuss the methods in \cite{Luscher:1996ug,Durr:2003nc}
to determine $\cA$, 
with emphasis on the one we finally used.

\subsection{Constant physics condition}

With two degenerate flavors of light quarks, the
theory has two bare parameters $\beta$ and $m_0$. The bare quark mass 
$m_0$ controls the physical quark mass and the bare coupling
determines the lattice spacing, defined
at vanishing quark mass (for a more precise discussion, which 
however is of little relevance in this context, see
\cite{Luscher:1996sc,Sommer:2003ne}).
Non--perturbative estimates of
\be
t_{r_0}(\beta)=\frac{[r_0/a](5.2)}{[r_0/a](\beta)}\;,
\ee
are available in a limited range of $\beta$ 
\cite{Gockeler:2004rp,Aoki:2002uc}.
In \cite{DellaMorte:2004bc}, the results of \cite{Gockeler:2004rp,Aoki:2002uc} 
were extrapolated
to zero quark mass. Taking directly these values for $[r_0/a](\beta)$,
we have the points with error bars in \fig{pt}.
From those we roughly estimated the location of the filled points,
using the perturbative dependence of the lattice spacing on $\beta$
as a guideline. For our action this is given to three loops in
\cite{Bode:2001uz}, which builds on various steps carried out in
\cite{Hasenfratz:1980kn,Kawai:1980ja,Weisz:1980pu,Sint:1995ch,Luscher:1995np,
Alles:1996cy,Christou:1998ws}. 
Applied as a pure expansion in the bare coupling (no tadpole improvement),
one has
\begin{eqnarray} \label{e:L3l}
{a(g_0^2) \over a((g_0')^2)} &=& e^{ -[g_0^{-2}-(g_0')^{-2}]/ 2b_0}
[g_0^2/(g_0')^2]^{-{b_1/2b_0^2}}\,
\left[\, 1 + q\, [g_0^2 -(g_0')^2] + {\rm O}\left((g_0')^4\right)\,\right], 
\\
&& q = 0.4529(1)\,,\quad g_0<g_0'\;. \nonumber
\end{eqnarray}
The evolution of the lattice spacing relative to our
reference point at  $(g_0')^2\!=\!6/5.2$ is then expressed by the function
$
t(\beta)=a(6/\beta)/a(6/5.2)$, 
which is plotted as a thick line in the graph. It confirms
that the filled points are very reasonable choices.
Note that other forms of applying bare perturbation theory 
(differing from \eq{e:L3l} in the $g_0^4$-term)
would give somewhat different results, but since we are 
interested in a rather limited range in
$g_0^2$ this does not matter much. 
Later we will show that
systematic uncertainties in $\cA$ introduced by this approximate scale
setting are negligible.
\EPSFIGURE[t]{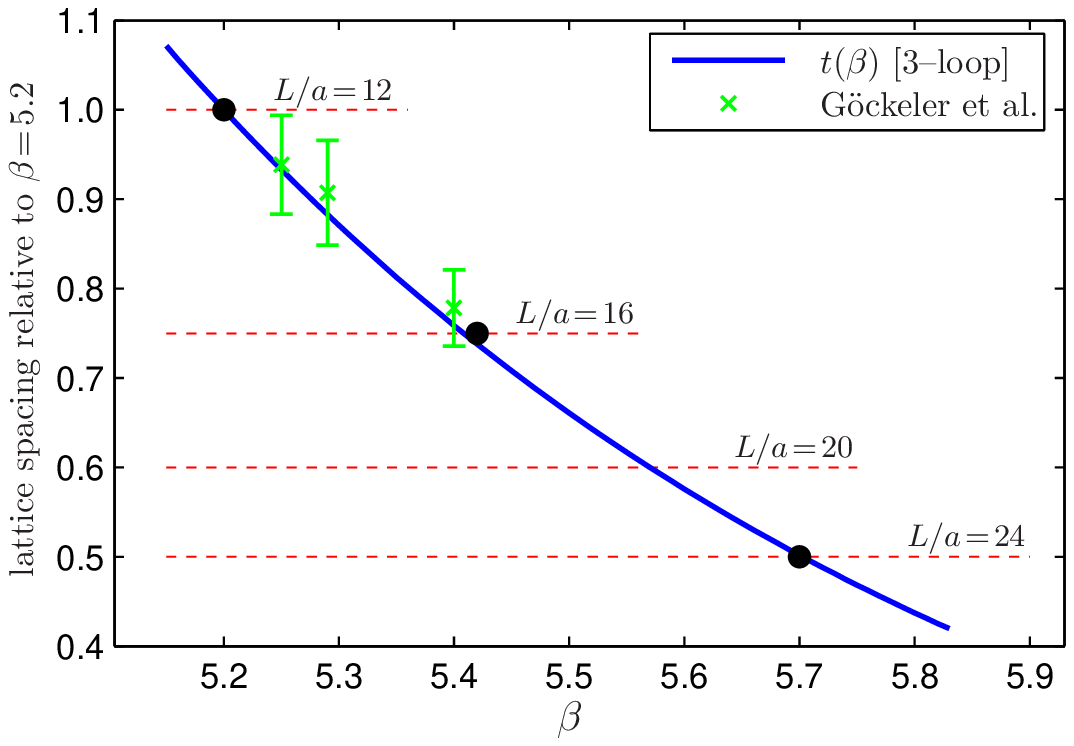,width=10cm}{
The evolution of the lattice spacing $a$ with the inverse bare
gauge coupling $\beta$ from perturbation theory in the lattice scheme
and large volume data \cite{Gockeler:2004rp} for the scale $r_0$.
The filled points correspond to our ``scaled'' simulations.
\label{pt}}

Finally, we keep the PCAC mass approximately constant.
A precise definition will be given in the following.

\subsection{Improvement conditions for the axial current}

In this section we discuss criteria for the choice of the improvement
condition.
With the isovector axial current
$A_\mu^a(x)=\psibar(x)\half\tau^a\dirac\mu\dirac5\psi(x)$ and the
corresponding pseudo--scalar
density $P^a(x)=\psibar(x)\half\tau^a\dirac5\psi(x)$ we define the
improved axial current \cite{Luscher:1996sc}
\be
(A_{\rm I})_\mu^a(x)=A_\mu^a(x)+a\cA\half(\partial_\mu+\partial^*_\mu)P^a(x)\;,
\label{impr}
\ee
where $\partial_\mu$ ($\partial_\mu^*$) denotes the forward (backward)
lattice derivative.
In the following we will consider quark masses derived from the PCAC relation
\be
m(x;\alpha,\beta)=\frac{\langle\alpha|\half(\partial_\mu+\partial^*_\mu)
(A_{\rm I})_\mu^a(x)|\beta\rangle}
{2\langle\alpha|P^a(x)|\beta\rangle}\;.
\ee
Since this mass is obtained from an operator identity, it is independent of
the states $|\alpha\rangle$ and $|\beta\rangle$ as well as the insertion point
$x$ up to cutoff--effects. Enforcing this independence at \emph{finite} lattice
spacing leads to possible definitions of improvement conditions
\cite{Durr:2003nc}. Inserting the
expression for the improved current (\ref{impr}) in the previous
equation, the
quark mass can be written as $m=r+a\cA s+\rmO(a^2)$ with
\bea
r(x;\alpha,\beta)&=&\frac{\langle\alpha|\half(\partial_\mu+\partial^*_\mu)
A_\mu^a(x)|\beta\rangle}
{2\langle\alpha|P^a(x)|\beta\rangle}\label{rgeneral}\\
\textrm{and}\qquad s(x;\alpha,\beta)&=&
\frac{\langle\alpha|\partial_\mu\partial^*_\mu P^a(x)|\beta\rangle}
{2\langle\alpha|P^a(x)|\beta\rangle}\;.\label{sgeneral}
\eea
If we now consider two sets of external states and two insertion points, the
improvement condition
$m(x;\alpha,\beta)=m(y;\gamma,\delta)$
yields
\be
-\cA=\frac{\Delta r}{a\Delta s}=\frac1a\!\cdot\!\frac{r(x;\alpha,\beta)
-r(y;\gamma,\delta)}
{s(x;\alpha,\beta)-s(y;\gamma,\delta)}
\ee
and therefore the sensitivity to $\cA$ is given by $a\Delta s$.

Once a reasonably large sensitivity is achieved, 
all improvement conditions at constant physics
are equally valid in the sense
that $\rmO(a)$ effects are removed in on-shell quantities. However, the way
in which \emph{higher--order} lattice artifacts are modified will depend on
the concrete choice of the improvement condition. In particular, if states with
energy not so far from the cutoff are involved, large $\rmO(a^2)$ effects might be
introduced.

We now specialize to Schr\"odinger functional boundary conditions, introduced in
\cite{Luscher:1992an,Sint:1993un} and recall the definition
of the relevant correlation functions. One considers  QCD in a finite
volume $L^3\times T$ with Dirichlet boundary conditions in time and periodic
boundary conditions in space. More precisely, the fermionic fields are periodic
up to a phase $\theta$. By taking functional derivatives with respect to
fermionic boundary source fields one can define correlation functions involving
the quark fields $\zeta,\zetabar$ at $x_0\!=\!0$ and $\zeta',\zetabar'$ at
$x_0\!=\!T$. In this work we use
\bea 
 \fa(x_0;\omega)&=&- \frac{a^3}{3L^6} \sum_\vecx\langle A_0^a(x) \, \op^a(\omega)
\rangle\;,\\
 \fp(x_0;\omega)&=&- \frac{a^3}{3L^6} \sum_\vecx\langle   P^a(x) \, \op^a(\omega)
\rangle\\ 
\textrm{and }\ \     f_1(
\omega',\omega)&=&- \frac1{3L^6}           \langle   \op'^a (\omega') \,
\op^a(\omega) \rangle
\eea
with the pseudo--scalar operator
\be
        \quad \op^a(\omega) =a^6 
         \sum_{\vecx,\vecy} \zetabar(\vecx) \gamma_5 \tau^a\half 
        \omega(\vecx-\vecy) \zeta(\vecy) \label{op}
\ee
at the $x_0\!=\!0$ boundary and the corresponding operator $\op'^a (\omega')$ at
the upper boundary of the SF cylinder. These operators depend on spatial 
trial ``wave functions'' $\omega$ and $\omega'$, respectively.

The Schr\"odinger Functional version of eqs. (\ref{rgeneral}) and (\ref{sgeneral}) 
is then given by
\bea
r(x_0;\omega)&=&\frac{\half(\partial_0+\partial^*_0)\fa(x_0;\omega)}
{2\fp(x_0;\omega)}\label{rSF}\\
\textrm{and}\qquad s(x_0;\omega)&=&\frac{\partial_0\partial^*_0 \fp(x_0;\omega)}
{2\fp(x_0;\omega)}\;.\label{sSF}
\eea
To determine $\cA$ in the $\nf\!=\!0$ theory \cite{Luscher:1996ug},
$\Delta r$ and $\Delta s$ were originally defined through a variation of the
periodicity angle $\theta$ of the fermion fields, while keeping $x_0\!=\!T/2$ and
$\omega\!=\!const$ fixed. For this method the sensitivity 
$a\Delta s$ is quite low when $L\gtrsim 0.8$fm, $T=2L$. In addition,
with dynamical fermions different values of $\theta$ would require separate
simulations. We therefore consider this method
as too expensive and disregard it in the following.
In the quenched approximation two alternatives have been explored in
\cite{Durr:2003nc}.

Requiring the quark mass to be independent of $x_0$ (for fixed $\theta$ and
$\omega\!=\!const$) is technically easy to implement. However, also in this case
the sensitivity is small unless large values of $\theta$ are used. Moreover,
the contribution of excited states is not well controlled, because 
one insertion point must be rather close to a boundary to achieve a
sufficiently large sensitivity. Thus energies which are not far
removed from the cutoff may contribute.

Secondly, variations of the wave function $\omega$ have been
considered. Ideally, one would like to use two wave functions
$\omega_{\pi^{(0)}}$ and $\omega_{\pi^{(1)}}$, such that the corresponding
operator $\op^a(\omega)$ couples only to the ground and first excited
state in the pseudo--scalar channel, respectively. As one can easily
see from \eq{sSF} the sensitivity to $\cA$ is then proportional to
\linebreak $m_{\pi^{(1)}}^2-m_{\pi^{(0)}}^2$. Higher excited states are (by definition) not
contributing and in principle the method can be used for rather small $T$.
Hence, we find this the most attractive method both from a theoretical and
practical point of view. In the next section we will detail our approximation
to this ideal situation.

\subsection{Wave functions}

We will now proceed to the more technical aspects of our
method. In order to approximate  $\omega_{\pi^{(0)}}$ and $\omega_{\pi^{(1)}}$
consider
a set of $N$ wave functions. Given a vector $u$ in this $N$--dimensional
space, projected correlation functions are defined as $(u,\fa)$ and $(u,f_1u)$,
i.e. $\fa$ is regarded as a vector and $f_1$ as a matrix in this space. It is
useful to represent $f_{\rm X}$ ($\rm X=A,P$) and $f_1$ as
\cite{Guagnelli:1999zf}
\bea
f_{\rm X}(x_0;\omega_i)&=&\sum_{n=0}^{M-1} F_{\rm X}^{(n)} v_i^{(n)}
e^{-m_\pi^{(n)}x_0}
+\rmO(e^{-m_\pi^{(M)}x_0})  +\rmO(e^{-m_\mrm{G}(T-x_0)})\;,\\
f_1(\omega_i,\omega_j)&=&\sum_{n=0}^{M-1} v_i^{(n)} v_j^{(n)}
e^{-m_\pi^{(n)}T} 
+\rmO(e^{-m_\pi^{(M)}T})+ \rmO(e^{-m_\mrm{G}T})\;,\label{f1decomp}
\eea
where $n$ labels the states in the pseudo--scalar channel in increasing energy
and $v_i^{(n)}$ is the overlap of such a state with the one generated by the
action of $\op^a(\omega_i)$ on the SF boundary state. The mass $m_\mrm{G}$ 
belongs to the lowest excitation in the scalar channel, the $0^{++}$
glueball and the coefficients $F_{\rm X}^{(n)}$
are proportional to the decay constant of the $n$th state.
Here we have suppressed the
explicit volume dependence of all quantities.

Knowledge of $v^{(n)}$ would allow the construction of vectors $u^{(n)}$,
such that -- up to corrections of order $e^{-m_\pi^{(M)}T}$ --
the correlation
$(u^{(n)},\fa)$ receives contribution from the $n$th state only. 
These $u^{(n)}$ may be computed
from the $v^{(n)}$ by a 
Gram-Schmidt orthonormalization. Clearly,
$u^{(0)}$ and $u^{(1)}$ can then be used to approximate $\omega_{\pi^{(0)}}$
and $\omega_{\pi^{(1)}}$.

An approximation to the $v^{(n)}$ 
can be obtained from the eigenvectors of the positive symmetric
matrix $f_1$. For the normalized eigenvectors
$\eta^{(0)},\eta^{(1)},\ldots$ corresponding to eigenvalues
$\lambda^{(0)}>\lambda^{(1)}>\ldots$ \eq{f1decomp} implies that
\bea
||\hat v^{(0)}-\eta^{(0)}||^2&=&\rmO(e^{-(m_\pi^{(1)}-m_\pi^{(0)})T})\\
\textrm{and}\quad
(\eta^{(1)},\hat v^{(0)})&=&\rmO(e^{-(m_\pi^{(1)}-m_\pi^{(0)})T})\;.
\eea
Thus, to the order indicated above, $\hat v^{(0)}$ is given by $\eta^{(0)}$
and $\eta^{(1)}$ is orthogonal to the ''ground state vector''
$\hat v^{(0)}$. As eigenvectors of a symmetric matrix the
$\eta^{(n)}$ are already orthogonal and we therefore use the approximation
\bea
\omega_{\pi^{(0)}}\simeq\sum_i\eta_i^{(0)}\omega_i &\quad \textrm{and}\quad &
\omega_{\pi^{(1)}}\simeq\sum_i\eta_i^{(1)}\omega_i\label{project}
\eea
to obtain correlators, which are (for intermediate $x_0$) dominated by the ground
and first excited state, respectively.
We note in passing that the ratios $v^{(n)}_i/v^{(n)}_j$ have a continuum limit if the
wave functions are properly scaled with the lattice spacing. 

In our simulations we restrict ourselves to a basis consisting
of three (spatially periodic) wave functions defined by
\bea
\omega_i(\vecx) &=& N_i^{-1} \sum_{{\bf n}\in{\bf Z}^3}
\overline{\omega}_i(|\vecx-{\bf n}L|)\,,\; i=1,\ldots,3\,,
\nonumber \\
\overline{\omega}_1(r) &=& r_0^{-3/2}\,\rme^{-r/a_0}\,,\quad \nonumber 
\overline{\omega}_2(r) =  r_0^{-5/2}\,r\,\rme^{-r/a_0} \,,\quad\\
\overline{\omega}_3(r) &=& r_0^{-3/2}\,\rme^{-r/(2a_0)}\,\;,
\eea
where $a_0$ is some physical length scale. We thus keep
it fixed in units of $L$,  choosing $a_0\!=\!L/6$.
The (dimensionless) coefficients $N_i$ are fixed to normalize the wave function via
$a^3\sum_{\vecx} \omega_i^2({\bf x})=1$. In addition we also consider the flat
wave function $\omega_0(\vecx)=L^{-3/2}$, where both quarks are projected to
zero momentum separately.\footnote{Since in this case $\vecx$ and $\vecy$ in
\eq{op} are uncorrelated, full translational invariance can be used without
performing additional inversions of the Dirac operator. 
For $\omega_{1\ldots3}$ we replace one of
the spatial sums in \eq{op} by a sum over 
eight far separated points, which means that one
performs eight times as many inversions. In a dynamical
fermion computation this additional effort is still 
small compared to the effort invested into the 
``updating''.}

\section{Numerical computation}

\subsection{Results for the improvement coefficient}

All our simulations were performed using non--perturbatively improved Wilson
fermions \cite{Luscher:1996sc,Luscher:1996ug,Jansen:1998mx,Yamada:2004ja} and
the plaquette gauge action.
For the boundary--improvement
coefficients $c_t$ and $\tilde c_t$ we used the 2--loop \cite{Bode:1999sm}
and 1--loop \cite{Sint:1997jx} values, respectively.
Concerning the algorithm, we employed the Hybrid Monte Carlo with two
pseudo--fermion fields as proposed in \cite{Hasenbusch:2001ne}. For all
observables we have checked the expected scaling of the statistical error
with the sample size and thus verified the absence of the problems described in
\cite{DellaMorte:2004hs} at the volumes and masses we consider here.
\tab{t_simpar} summarizes the parameters of our simulations.

\TABULAR[!h]{r|rrccrll}{
    \hline
 run & $L/a$ &  $T/a$  &  $\beta$ & $\kappa$ & $N_{\rm meas}$  &  $\ am/t(\beta)$ &  
$\quad-\cA$ \\
    \hline
\rm I   & 12  & 12  & 5.20 & 0.135600 & $320\ $&  $0.0151(9)$ & $0.0638(23)$ \\
\rm II  & 16  & 16  & 5.42 & 0.136300 & $200\ $&  $0.0171(5)$ & $0.0420(21)$ \\
\rm III & 24  & 24  & 5.70 & 0.136490 & $120\ $&  $0.0151(4)$ & $0.0243(36)$ \\\hline
\rm IV  & 12  & 12  & 5.20 & 0.135050 & $160\ $&  $0.0363(6)$ & $0.0697(31)$ \\
\rm V  & 16  & 20  & 5.57 & 0.136496  & $290\ $&  $0.0154(4)$ & $0.0366(36)$ \\
\rm VI & 24  & 24  & 6.12 & 0.136139  & $40\ $ &  $0.0002(4)$ & $0.0244(21)$ \\\hline}
{Summary of simulation parameters and results for $\cA$. 
Runs {\rm I-III} are at constant physics.\label{t_simpar}}

The $\beta$ values for run {\rm II} and {\rm III} have been chosen such
that $L/r_0$ is approximately the same as in
run {\rm I}, which corresponds to $L\simeq1.2\fm$. In exploratory
quenched studies \cite{Durr:2003nc} this volume was found to be sufficient for
the described projection method to work. 
$N_{\rm meas}$ is the number of  estimates
of $\cA$, separated by 4--12 unit length HMC trajecrories. 
The autocorrelation of these measurements turned out 
to be negligible.
The column labeled $am/t(\beta)$ refers to the
bare quark mass $m=r(T/2;\omega_0)+a\cA s(T/2;\omega_0)$, cf.
eqs.~(\ref{rSF}, \ref{sSF}). The 1--loop value of $\cA$ from
\cite{Luscher:1996vw}
is used there. We tuned the hopping parameter $\kappa$ in order to keep $am/t(\beta)$
fixed when varying $\beta$, thus ignoring (presumably small) changes in the
renormalization factors. 
Note that we have chosen a finite, but small bare quark mass
of around 30~MeV. Such a mass helps (in addition to the Dirichlet
boundary conditions) to reduce the cost of the simulations.
Results from the remaining simulations are used to
discuss systematic uncertainties in our determination of $\cA$.

\DOUBLEFIGURE[t]{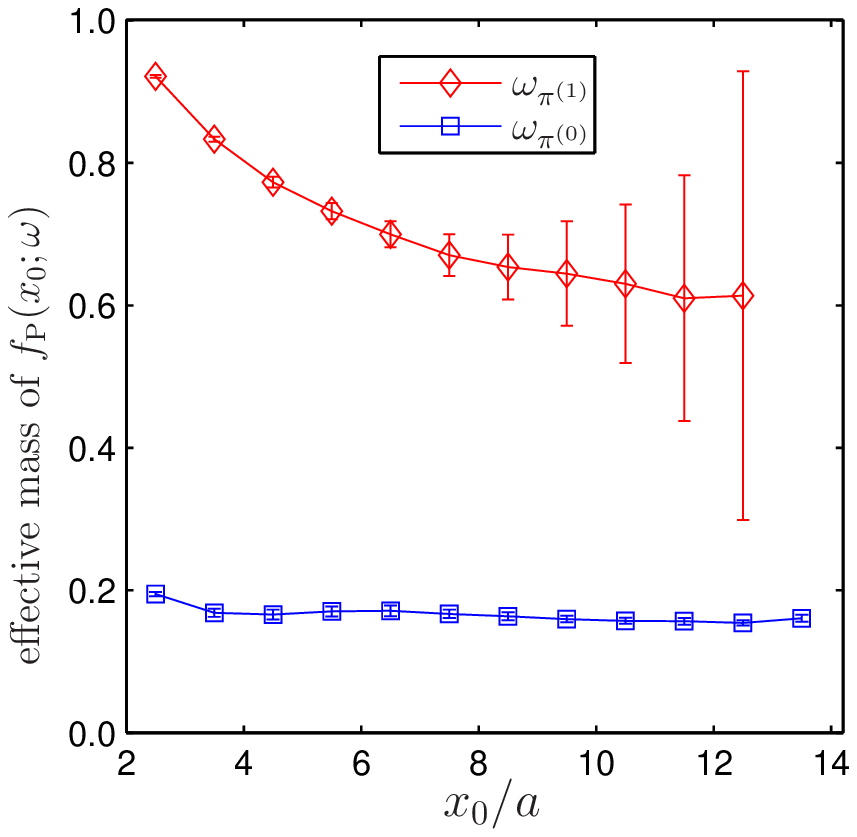,height=6.55cm}{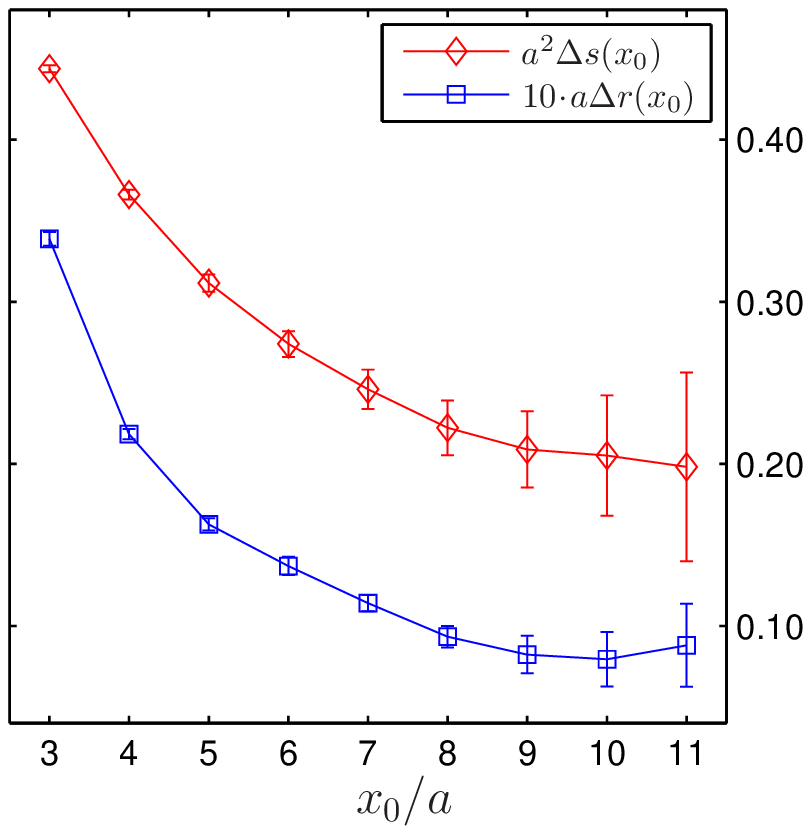,height=6.45cm}{
The effective mass in lattice units 
of the projected correlation functions $(\eta^{(0)},\fp)$
and $(\eta^{(1)},\fp)$ from run {\rm II}.
\label{effmass}}
{$\Delta s(x_0)$ and $\Delta r(x_0)$ determined from $\omega_{\pi^{(0)}}$ and
$\omega_{\pi^{(1)}}$ in run {\rm II}.
\label{effrs}}

In \fig{effmass} we show the effective masses from $\fp(x_0;\omega_{\pi^{(0)}})$
and $\fp(x_0;\omega_{\pi^{(1)}})$ as obtained in run {\rm II}.
Two distinct signals are clearly visible, which indicates that the
described approximate projection method works well at these parameters.
The energy of the first excited state is not far away from $a^{-1}$,
suggesting that in even smaller volumes the residual $\rmO(a^2)$ effects would
grow rapidly.
In the spirit of the remark after \eq{project} at the other values of $\beta$ we used
the same linear combination of wave functions to define $\omega_{\pi^{(0)}}$ and
$\omega_{\pi^{(1)}}$, namely
\be
\begin{array}{rcl}
\eta^{(0)}&=&(\,0.5172,\,\phantom{-}0.6023,\,\phantom{-}0.6081\,)\\
\textrm{and}\quad
\eta^{(1)}&=&(\,0.8545,\,-0.3233,\,-0.4066\,)\;,\label{eta}
\end{array}
\ee
which are the ones determined in run {\rm II}.
When scaled in units of $r_0$, this yields effective masses similar to those 
shown in \fig{effmass}.
Results from a redetermination of $\eta^{(0)}$ and $\eta^{(1)}$ in the other
matched simulations would differ from \eq{eta} only by $\sim1\%$.
In the figure the error on the effective mass of the first excited state is
seen to be quite large,
but what actually enters the computation of $\cA$ is the error of
\bea
\Delta r(x_0)=r(x_0;\omega_{\pi^{(1)}})-r(x_0;\omega_{\pi^{(0)}})
&\ \textrm{ and }\ &\Delta s(x_0)=s(x_0;\omega_{\pi^{(1)}})
-s(x_0;\omega_{\pi^{(0)}})
\;.
\eea
These profit from statistical correlations of the correlation functions
entering their definition and thus have smaller statistical
errors as can be seen
in \fig{effrs}, where we plot $a\Delta r$ and $a^2\Delta s$ from the same data
used in \fig{effmass}.

\fig{effca} collects results
for the ''effective'' $\cA(x_0)=-\Delta r(x_0)/a\Delta s(x_0)$
from the matched runs {\rm I-III}.
We see little variation for $x_0\gtrsim 6a$, which we take as another signal 
that high energy states which could contribute large 
$\rmO(a)$ ambiguities in the improvement condition
are reasonably suppressed in this region.
We complete
our definition of $\cA$ with the choice $x_0\!=T/2$,
which is at the same time scaled in physical units and in agreement
with the $x_0\gtrsim 6a$ bound for all our lattices.

\EPSFIGURE[l]{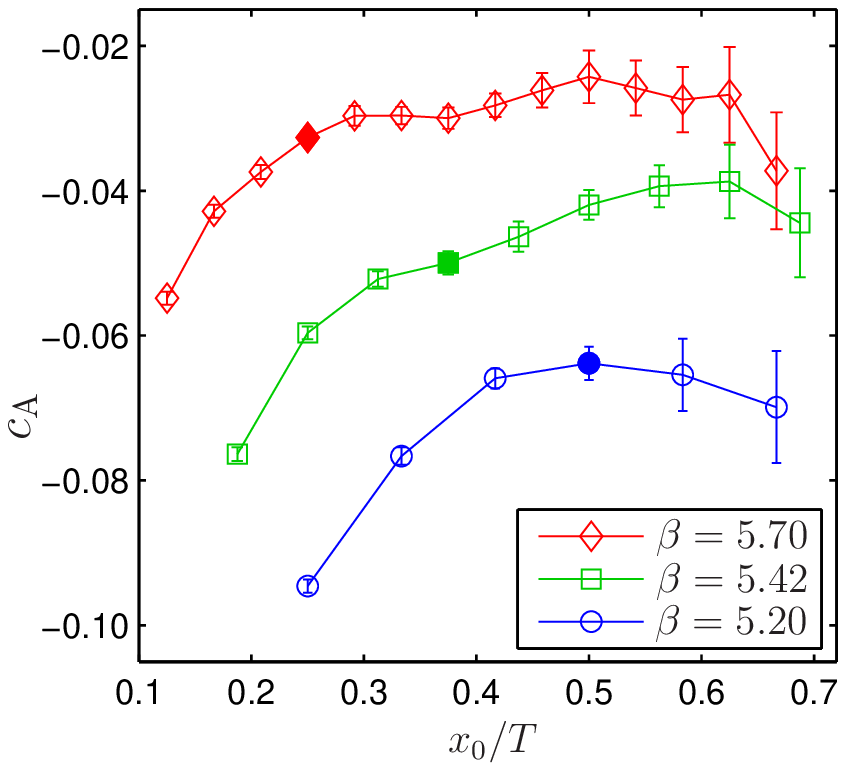,width=67mm}{
Effective $\cA$ as determined from $\omega_{\pi^{(0)}}$ and
$\omega_{\pi^{(1)}}$ for runs {\rm I-III}. Points 
with $x_0/a=6$ are marked 
by filled symbols.
\label{effca}}

Finally, $\cA$ is plotted as a function of $g_0^2$ in \fig{figpade}. 
The solid line is a smooth interpolation of the data from
the matched simulations, constrained in addition by 1--loop perturbation
theory:
\begin{equation}
\cA(g_0^2)=-0.00756\, g_0^2\times\frac{1-0.4485\, g_0^2}{1-0.8098\, g_0^2}.
\end{equation}
It is our final result, valid in the range $0.98 \leq g_0^2 \leq1.16$  
within the errors of the data points (at most 0.004). 
\EPSFIGURE[b]{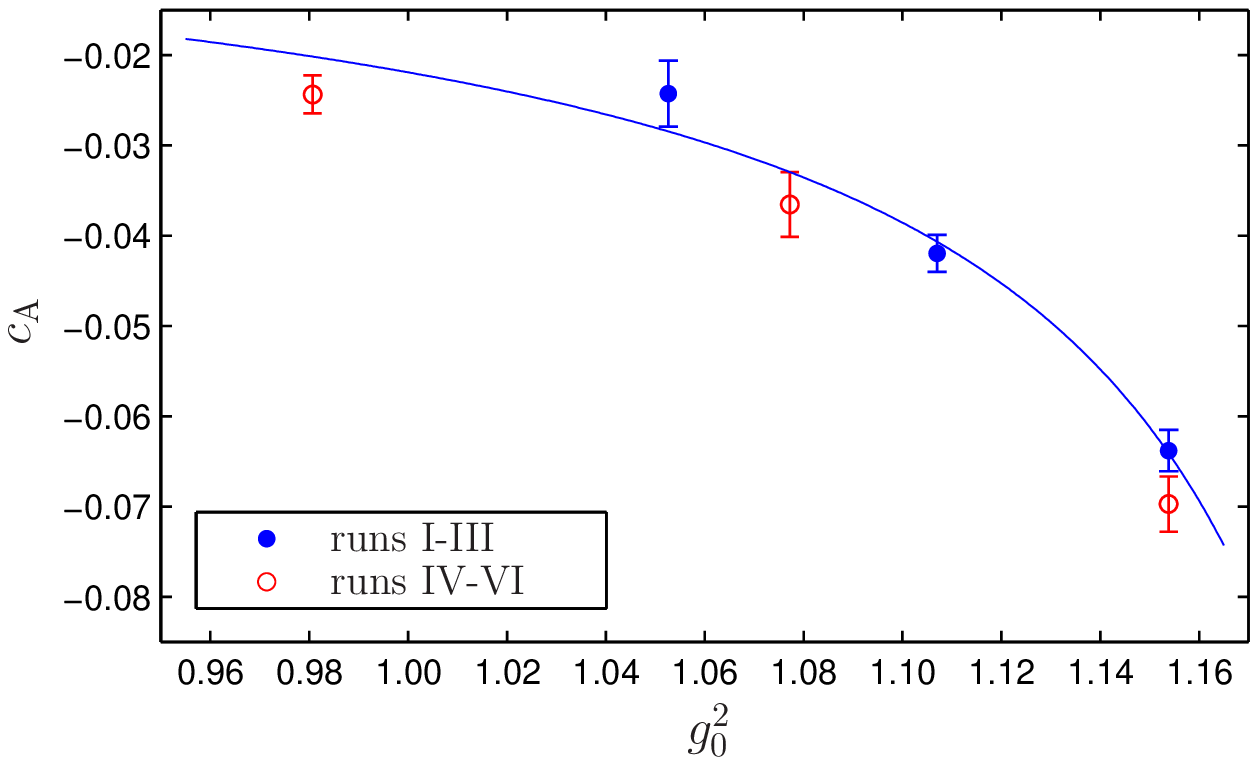,width=12cm}{
Simulation results for $\cA$. The solid line represents a fit
of the data points at constant physics (filled circles).
\label{figpade}}

The non-perturbative result is quite far away form 
1-loop perturbation theory, which takes
a value of $\cA=-0.0087$ at $g_0^2=1.15$.
Using data from \cite{Irving:2001vy}, we see that the effect
on the result for the pseudo--scalar decay constant at this lattice spacing
is as large as $10\%$.

\subsection{Uncertainties due to deviations from the ``constant physics''
condition}

We should check whether the volumes in our runs {\rm I-III} are scaled
sufficiently precisely or if systematic errors need to
be added to the statistical ones on $\cA$ to cover possible
violations of the constant physics condition.
Table 1 shows that the bare PCAC mass has been
kept constant to within about 10\%. A renormalized quark
mass differs by the multiplication with a $Z$-factor which is
a slowly varying function of $a$. As explained in the beginning of sect.~2,
such a factor is irrelevant. Run {\rm IV} is done with a quark mass which is more
than a factor 2 larger than the one in run {\rm I}, with otherwise identical
parameters. The small difference in $\cA$ confirms that the small deviations
from the ``constant mass'' condition can be neglected.

The other issue is the uncertainty due to our perturbative (or asymptotic)
scaling of the physical length scales. It has been argued in sect.~2.1 that
the difference to a proper non-perturbative scaling is rather small.
Also, estimating a possible change by comparing 3-loop to 2--loop and
non--perturbative scaling,
gives a deviation in $t(\beta)$ which is smaller than 10\% in
the whole range of \fig{pt} and thus the
same maximum deviation applies to $L/a$. Again this can be neglected altogether.
Additional confirmation comes from a comparison of
the result from run {\rm V} to our fit formula. In run  {\rm V},
$L/a$ is 20\% lower than the proper value, but $\cA$
does not differ significantly from the fit curve.

Finally, by run {\rm VI} we verify that the dependence of $\cA$ on the
kinematic parameters disappears quickly when going to even larger values of $\beta$.
In this run we used gauge configurations from the calculation of $Z_{\rm P}$
\cite{DellaMorte:2003jj}. Although those were produced at $m\!=\!0$,
$\theta\!=\!0.5$ and
a much smaller volume, the resulting $\cA$ is only approximately two standard
deviations away from our fit.

\section{Discussion}

For the $\rmO(a)$-improved action with non-perturbative $\csw$
\cite{Jansen:1998mx},
we have determined the improvement coefficient $\ca$ 
for $\beta\!\geq\!5.2$, which roughly corresponds to $a\!\leq\!0.1$~fm. 
The improvement condition was implemented at constant physics, which is 
necessary in the situation when $\rmO(a)$ ambiguities in the improvement
coefficients are not negligible. 
This is indeed the case here: at $\beta\!=\!5.42$ we have evaluated
$\ca$ also for an $L/a\!=\!12, T/a\!=\!16$ geometry. The value of $\ca$ is about 
a (statistically significant) 30\% larger in magnitude than for the constant physics
condition $L/a\!=\!16\!=\!T/a$. Thus, imposing the constant physics condition is
important in the present case. 

In addition, in order
to safely exclude large $\rmO(a)$ ambiguities, improvement conditions
should only involve states with energy $E\ll a^{-1}$. On this 
requirement we had to compromise more than we would have liked to do.
Our maximum values for $Ea$ are about $0.7$. Although this
could have been improved
by going to somewhat larger values of $L$ (and $T$), this would
have made the numerical computation much more expensive. 

We finally note that large $\rmO(a^2)$ effects have been found in the $\nf\!=\!2$,
$\rmO(a)$-improved theory \cite{Sommer:2003ne} at $\beta\!=\!5.2$. These may well
be related to the not so small $\rmO(a)$ ambiguity in $\ca$
that we just mentioned. This can only be investigated further by 
studying  the scaling violations
in quantities such as $F_{\pi}r_0$ after improvement. Of course also 
the renormalization constant $\za$ has to be known to carry out 
such a study. We are presently
computing  $\za$ following the strategies of \cite{Luscher:1996jn,Hoffmann:2003mm}.
As an immediate application one can then $\rmO(a)$--improve
and renormalize the bare pseudoscalar decay constants
computed in \cite{Allton:2001sk,Irving:2001vy,Aoki:2002uc}. 

Clearly, the method employed in this paper may also be useful to
compute $\ca$ in the three flavor case, where $\csw$ is known
non--perturbatively with plaquette and Iwasaki gauge actions 
\cite{Aoki:2002vh,Yamada:2004ja,Ishikawa:2003ri}.


\bigskip

\acknowledgments

We thank Stephan D\"urr and Heiko Molke for their valuable contributions in the
early stages of this project.
Discussions with S.~Aoki, S.~Hashimoto and T.~Kaneko and comments from
T.~Kaneko and U.~Wolff on the manuscript are gratefully acknowledged. 
We thank NIC/DESY Zeuthen for allocating computer time on the
APEmille machines for this study.
This work is supported in part by the Deutsche Forschungsgemeinschaft
in the SFB/TR 09-03, ``Computational Particle Physics'' and the
Graduiertenkolleg GK271.

\newpage
\appendix

\bibliography{refs}           
\bibliographystyle{JHEP}   

\end{document}